\documentclass{elsart}
\usepackage{epsfig}
\usepackage{graphicx}
\usepackage{amsmath}
\usepackage{amsfonts}
\usepackage{amssymb}%
\begin{document}
\title{The rheology of concentrated suspensions of arbitrarily-shaped particles }
\author{I. Santamar\'{\i}a-Holek$^{1}$ and Carlos I. Mendoza$^2$}
\address{$^{1}$Unidad Multidisciplinaria de Docencia e Investigaci\'{o}n, Facultad de
Ciencias Juriquilla, Universidad Nacional Aut\'{o}noma de M\'{e}xico, Boulevard
Juriquilla 3001, Quer\'{e}taro, 76230, Mexico}
\address{$^{2}$Instituto de Investigaciones en Materiales, Universidad Nacional
Aut\'{o}noma de M\'{e}xico, Apdo. Postal 70-360, 04510 M\'{e}xico, D.F., Mexico}

\begin{abstract}
We propose an improved effective-medium theory to obtain the concentration
dependence of the viscosity of particle suspensions at arbitrary volume
fractions. Our methodology can be applied, in principle, to any particle shape
as long as the intrinsic viscosity is known in the dilute limit and the
particles are not too elongated. The procedure allows to construct a
continuum-medium model in which correlations between the
particles are introduced through an effective volume fraction. We have tested
the procedure using spheres, ellipsoids, cylinders, dumbells, and other
complex shapes. In the case of hard spherical particles, our expression
improves considerably previous models like the widely used
Krieger-Dougherty relation. The final expressions obtained for the viscosity
scale with the effective volume fraction and show remarkable agreement with
experiments and numerical simulations at a large variety of situations.
\end{abstract}
\maketitle


\section{Introduction}

When particles are suspended in an homogeneous isotropic fluid, the
viscosity of the resulting complex fluid is increased. In the case of dilute
suspensions, the increase in viscosity as a function of the volume fraction
$\phi$ was firstly determined for spherically shaped particles by Einstein in
1911, \cite{einstein}. Some years later, extensions of Einstein's work
appeared and new formulas for the viscosity as a function of the volume
fraction for solid ellipsoidal particles and emulsions were also derived by
Jeffery (1922) and Taylor (1932), respectively \cite{jeffery,taylor}.

For sufficiently low particle concentrations, the viscosity $\eta$ of a
suspension can in general, be written as,%
\begin{equation}
\eta\left(  \phi\right)  =\eta_{0}\left(  1+\left[  \eta\right]  \phi
+k_{H}\phi^{2}+...\right)  , \label{lowconcentration}%
\end{equation}
where $\eta_{0}$ is the solvent viscosity, $\left[  \eta\right]  $ is
the low filling fraction intrinsic viscosity, $k_{H}$ is the so
called Huggins coefficient, and $\phi$ is the volume fraction of the
particles. The value of $\left[  \eta\right]  $ depends on the
particle shape. Although its calculation is difficult, there are few cases
where $\left[  \eta\right]  $ has been obtained analytically as in the case of
spheres\cite{einstein}, ellipsoids \cite{rallison}, \cite{haber}, long
cylinders \cite{brenner} and dumbbells consisting of two identical spheres
\cite{brenner}, \cite{wakiya}. Fortunately, accurate numerical approaches have
been developed to calculate the intrinsic viscosity of arbitrarily shaped
particles \cite{zhou}, \cite{wierenga}.

The large amount of work devoted to determine the relation between viscosity
and concentration under different physical conditions that reflect
experimental protocols \cite{vandeven,larson,brennerLibro,doi}, is clearly
connected with the important role that suspensions and emulsions
play in almost all fields of industry, medicine and biology-related
soft-matter systems.

Many remarkable theoretical and, more recently, numerical works have made
central contributions to the understanding of the rheology of concentrated
suspensions (see for instance, Refs. \cite{saito}-\cite{russel}). Essentially,
most of these works introduce particle correlations by taking into account
hydrodynamic interactions and provide a conceptual framework that explains how
the microstructure of the systems modifies the behavior of the viscosity as a
function of the volume fraction. However, the quantitative description of the
problem still constitutes an open challenge for theoretical descriptions even
in the simplest case of spherical particles.

Several models have been proposed in order to extend the range of
applicability of Einstein's expression to larger volume fractions
\cite{vandeven}. Semi-empirical \cite{quemada,bicenario} and effective-medium
\cite{brinkman}-\cite{pal} models intended to extend the quantitative
description of the experimental dependence of the effective viscosity of hard
particle suspensions at arbitrary volume fractions have been proposed long
ago. Among them we can mention the one derived by Saito \cite{saito} that
accounts for hydrodynamic interactions between uncorrelated particles%
\begin{equation}
\eta\left(  \phi\right)  =\eta_{0}\left[  1+\left[  \eta\right]  \left(
\frac{\phi}{1-\phi}\right)  \right]  . \label{saito}%
\end{equation}

Other approaches include the use of differential effective-medium
theories (DEMT). Krieger and Dougherty \cite{krieger} derived the following
expression that incorporates excluded volume using a crowding effect
introduced by Mooney \cite{mooney}%
\begin{equation}
\eta\left(  \phi\right)  =\eta_{0}\left(  1-\frac{\phi}{\phi_{\max}}\right)
^{-\left[  \eta\right]  \phi_{\max}}, \label{krieger}%
\end{equation}
with $\phi_{\max}$ the filling fraction at maximum packing. This formula
agrees reasonably well with the experimental data particularly in the
low concentration regime. Moreover, this relation reduces to the correct
Einstein's equation in the limit of infinite dilution. Very recently,
Bullard and coworkers \cite{bullard} also used DEMT techniques to derive a
similar relationship for suspensions of particles that may themselves
incorporate some of the solvent either by solvation or by occlusion in
interstitial pores. In the case of spheres their expression adopts the form%
\begin{equation}
\eta\left(  \phi\right)  =\eta_{0}\left(  1-K\phi\right)  ^{-\left[
\eta\right]  /K}, \label{bullard1}%
\end{equation}
where the factor $K$ considers flocculation of the particles thus
representing the ratio of the volume of the clusters to the volume of the
particles forming the clusters \cite{bullard}.

The original formulations of DEMT were based on Einstein's expression,
providing moderate agreement with experimental results. It is important to
stress that one common characteristic of these models is that they describe
acceptably well the viscosity-concentration relation at low concentrations but
fail to describe correctly the whole concentration regime. As we will explain
later, this is because these theories do not incorporate appropriately the
correlations introduced by the excluded volume effects.

A different approach to calculate the viscosity of suspensions is based
on the expected theoretical divergence of viscosity near the percolation
threshold \cite{brady}%
\begin{equation}
\eta\left(  \phi\right)  \sim\eta_{0}\left(  1-\frac{\phi}{\phi_{c}}\right)
^{-2}, \label{quemada}%
\end{equation}
which, according to Douglas and coworkers\cite{douglas}, should be
independent of the shape of the particles in suspension.

Bicerano et al.\cite{bicenario} examined the viscosity of suspensions of
different hard bodies, and proposed a formula for the relative viscosity that
provides a smooth transition between the dilute and the concentrated,
Eq.(\ref{quemada}), regimes and is valid for low-shear%
\begin{equation}
\eta\left(  \phi\right)  =\eta_{0}\left(  1-\frac{\phi}{\phi_{c}}\right)
^{-2}\left[  1+C_{1}\left(  \frac{\phi}{\phi_{c}}\right)  +C_{2}\left(
\frac{\phi}{\phi_{c}}\right)  ^{2}\right]  ,\label{bicenario}%
\end{equation}
with%
\begin{equation}
C_{1}=\left[  \eta\right]  \phi_{c}-2,\label{c1}%
\end{equation}
and%
\begin{equation}
C_{2}=k_{H}\phi_{c}^{2}-2\left[  \eta\right]  \phi_{c}+1.\label{c2}%
\end{equation}
This theoretical formula accounts for both the low density exact results
from hydrodynamics as well as the large volume fraction semi-empirical
expansions. Pryamitsyn and Ganesan \cite{pryamitsyn} proposed to extend the
domain of validity of this expression to arbitrary shear rates by considering
the viscosity percolation threshold $\phi^{\ast}($\emph{Pe}$)$ a
function of the Peclet number.

Recently, a model has been proposed that introduces in appropriate form the
correlations introduced by the excluded volume effects and gives an excellent
quantitative description of the viscosity of solid and liquid suspensions of
spherical particles at arbitrary filling fractions Ref.
\cite{mendoza,mendoza2}. This model incorporates an effective filling fraction
$\phi_{eff}$ that leads to an universal representation of all experimental
results in a master curve that suggest the role of $\phi_{eff}$ as a scaling
variable for the viscosity of these systems.

Despite the importance of the theoretical and numerical analysis performed
mostly in the case of spherical particles, in real systems the assumption of
sphericity of the suspended particles is not always satisfied since
polydispersity and different particle shapes have to be taken into account
\cite{wagner}. Thus, the main objective of this article is to propose a
continuum-medium description for the viscosity-concentration relation for
different particle shapes as long as they are not too elongated. The
description is derived using DEMT techniques introducing correlations between
particles through an effective volume fraction that incorporates excluded
volume effects. These effects are responsible for the scaling properties of
the suspensions and show to be universal independently of the shear rate and
the shape of the particles.

The article is organized as follows, in Section II we propose a first
correction to Eq.(\ref{lowconcentration}) that takes into account excluded
volume effects and use it as the starting point of a differential effective
medium approach. In Section III we compare the predictions of our model with
previous theories and with various experimental results for different particle
shapes. Finally Section IV is devoted to conclusions.

\section{Correlations and DEMT approach}

The main difficulty when dealing with suspensions at large concentrations is
to take into account the correlations among particles. Although many efforts
have been devoted to microscopically calculate the viscosity of a suspension
of hard particles taking into account the hydrodynamic interactions, the
enormous mathematical complications associated to the many body problem only
permit to find corrections applicable to the low concentration regime.
However, in first approximation, such correlations can be considered as
follows: the contribution by the particles to the total stress tensor
$\overline{\mathbf{\Pi}}_{p}^{V}$ of the suspension is given in terms
of an average over the volume $V$ of the system in the form %
\begin{equation}
\overline{\mathbf{\Pi}}_{p}^{V}\simeq\frac{N}{V}\int\mathbf{\Pi}_{p}^{(1)}dV,
\label{landau22.5}%
\end{equation}
where we represented the single particle contribution to the stress
tensor by $\Pi_{p}^{(1)}$. The upper $V$ in Eq.(\ref{landau22.5}) 
stands for the volume average and the factor $N$ accounts for the
contribution of the $N$ independent particles. However, as defined in
Eq.(\ref{landau22.5}), the average is strictly valid only for a system of
point particles \cite{mendoza}.

Considering that each particle has a volume $V_{p}$, then the
volume average must be performed over the free volume accessible to the
particles defined by: $V_{free}=V-cNV_{p}$. Here $c$ is a
geometric factor that takes into account the fact that the complete free
volume cannot be filled with particles. Note that, for different shapes of the
suspended particles, the value of the constant $c$ will be different.
It also contains information about the maximum packing of particles the system
may allocate and if the excluded volume effects are taken into account, the
suspended phase contribution to the stress tensor is now given by
\cite{mendoza}, \cite{mendoza2}%
\begin{align}
\overline{\mathbf{\Pi}}_{p}^{V_{free}}  &  \simeq\frac{N}{V-cNV_{p}}%
\int\mathbf{\Pi}_{p}^{(1)}dV,\nonumber\\
&  =\frac{1}{1-c\phi}\left[  \eta\right]  \phi\overline{\nabla v_{0}^{0}}^{V},
\label{landau22.5-B}%
\end{align}
where $\phi=NV_{p}/V$, $\overline{\nabla v_{0}^{0}}^{V}$
is the volume average of the traceless velocity gradient with $v_{0}$
the velocity field of the background fluid \cite{landau}, \cite{mendoza}.
For finite-sized particles, this relation leads to the result that
Einstein's expression scales with the excluded volume factor $\phi/(1-c\phi
)$ instead of $\phi$, and thus gives the following expression for
the viscosity of a suspension%
\begin{equation}
\eta\left(  \phi\right)  =\eta_{0}\left(  1+\left[  \eta\right]  \phi
_{eff}\right)  , \label{lowconcentration+correlations}%
\end{equation}
where the effective filling fraction $\phi_{eff}$ is defined by%
\begin{equation}
\phi_{eff}=\frac{\phi}{1-c\phi}. \label{phieff}%
\end{equation}
The constant $c$ depends on the filling fraction $\phi_{c}$ which is the
critical concentration at which the suspension loses its fluidity. This
viscosity percolation threshold generally is greater than the purely
geometrical percolation threshold of the particles, but less than or equal to
the maximum packing fraction $\phi_{\max}$ \cite{bullard} and is given
by%
\begin{equation}
c=\frac{1-\phi_{c}}{\phi_{c}}. \label{c}%
\end{equation}
The effective filling fraction (\ref{phieff}) approaches the bare $\phi$ at
low concentrations and becomes $1$ at the divergence of the viscosity which
occurs at $\phi_{c}$. The fact that the particles can not occupy all the
volume of the sample due to geometrical restrictions is taken into account in
the crowding factor $c$. For example, for a face centered cubic (FCC)
arrangement of identical spheres, the maximum volume that the spheres may
occupy is larger than for a random arrangement of spheres.

The formula for the effective viscosity of the suspensions given by relation
Eq. (\ref{lowconcentration+correlations}) can also be written as%
\begin{equation}
\eta\left(  \phi\right)  =\eta_{0}\left[  1+\left[  \eta\right]  \left(
\frac{\phi}{1-c\phi}\right)  \right]  . \label{lowconcentration+correlations2}%
\end{equation}
Although in this expression it is clear that $\eta\left(  \phi\right)  $
incorporates the excluded volume corrections for the viscosity of a
suspension, hydrodynamic interactions are ignored and therefore one expects
its validity been restricted to low concentrations. To improve it, further
corrections must appear due to the interactions between particles. In the
system under consideration, these interactions are the hydrodynamic
interactions which become increasingly important when increasing the filling
fraction. The mentioned correlations can be accounted for by using DEMT
techniques \cite{bullard}. This theoretical method is based on a progressive
addition of spheres to the sample in which the new particles interact in an
effective way with those added in previous stages \cite{vandeven}.

Taking Eq. (\ref{lowconcentration+correlations}) as the starting point,
suppose that we increase by $\delta\phi_{eff}$ the particle concentration in
the suspension of viscosity $\eta\left(  \phi_{eff}\right)  $ by adding a
small quantity $\Delta\phi_{eff}$ of few new particles. If we treat the
suspension into which we add these particles as a homogeneous effective
medium of viscosity $\eta\left(  \phi_{eff}\right)  $, then the new viscosity
can be written as%
\begin{equation}
\eta\left(  \phi_{eff}+\delta\phi_{eff}\right)  =\eta\left(  \phi
_{eff}\right)  \left(  1+\left[  \eta\right]  \Delta\phi_{eff}\right)  .
\label{recursive1}%
\end{equation}
Note that the increase in the effective particle concentration %
$\delta\phi_{eff}$ is different from the effective concentration of new
particles added at a given stage $\Delta\phi_{eff}$. This is due to the
fact that one has to remove part of the effective medium, which already
contains some particles, in order to allocate the new particles. From this, it
follows that the fraction of particles of the new effective medium is given by
$\phi_{eff}+\delta\phi_{eff}=\phi_{eff}(1-\Delta\phi_{eff})+\Delta\phi_{eff}%
$, from which we find%
\begin{equation}
\Delta\phi_{eff}=\frac{\delta\phi_{eff}}{1-\phi_{eff}}. \label{deltaphi}%
\end{equation}
Substituting Eq.(\ref{deltaphi}) in Eq.(\ref{recursive1}) and
integrating we finally obtain%
\begin{equation}
\eta\left(  \phi\right)  =\eta_{0}\left(  1-\phi_{eff}\right)  ^{-\left[
\eta\right]  }, \label{viscosity2}%
\end{equation}
or, using the definition of $\phi_{eff}$%
\begin{equation}
\eta\left(  \phi\right)  =\eta_{0}\left[  1-\left(  \frac{\phi}{1-c\phi
}\right)  \right]  ^{-\left[  \eta\right]  }. \label{viscosity1}%
\end{equation}
This relation for the effective viscosity of a suspension of rigid particles
constitutes a powerful improved generalization of previous theoretical results
and empirical proposals, as we will show in the following section. In the
limit of low concentrations Eq.(\ref{viscosity1}) always reduces to
Eq.(\ref{lowconcentration}) and the predicted Huggings coefficient,
obtained by expanding Eq.(\ref{viscosity1}) in a virial series, is given by
\begin{equation}
k_{H}=\frac{1}{2}\left[  \eta\right]  \left(  \left[  \eta\right]  +\frac
{2}{\phi_{c}}-1\right)  , \label{kH}%
\end{equation}
that has the form suggested by Douglas (see Eq.(17) of
Ref.\cite{bullard}).

Although the procedure that leads to Eq. (\ref{viscosity1}) is similar
to other DEMT \cite{brinkman}-\cite{pal},\cite{bullard}, the introduction of
excluded volume correlations through the effective filling fraction 
$\phi_{eff}$ defined in Eq.(\ref{phieff}) and then its role as
integration variable in the differential procedure improves remarkably the
agreement with experimental data when compared to other models, this is shown
in Ref.\cite{mendoza} for the special case of hard spheres. This agreement is
due to the fact that the use of $\phi_{eff}$ as integration variable
implicitly considers correlations between spheres of the same recursive stage
in contrast to other models.

It will be shown in the next sections that our model gives excellent
quantitative results at the whole concentration range for different particle
shapes. However, at this point it is interesting to compare the functional
form of our expression (\ref{viscosity1}) with the one by Bicerano and
coworkers, Eq.(\ref{bicenario}). Their equation was designed explicitly to
provide a smooth transition between the semidilute and concentrated regimes.
Equation (\ref{viscosity1}) can be written in a form similar to Bicerano's
expression to obtain%
\begin{equation}
\frac{\eta\left(  \phi\right)}{\eta_{0}}  =\left(  1-\frac{\phi}{\phi_{c}}\right)
^{-2}\left[  1+C_{1}\left(  \frac{\phi}{\phi_{c}}\right)  +C_{2}\left(
\frac{\phi}{\phi_{c}}\right)  ^{2}+C_{3}\left(  \frac{\phi}{\phi_{c}}\right)
^{3}+O\left(  \phi^{4}\right)  \right]  .\label{viscosity1-bicenario}%
\end{equation}
Here, $\phi_{c}$ can be considered a function of the P\'{e}clet
number and therefore is no restricted to the low-shear regime. The constants
$C_{1}$ and $C_{2}$ are identical to the ones corresponding to
the model of Bicerano and coworkers given by Eqs.(\ref{c1}) and (\ref{c2}),
and the coefficient of the cubic term is given by%
\begin{equation}
C_{3}=\frac{\phi_{c}^{3}}{6}\left[  \eta\right]  \left(  \left[  \eta\right]
^{2}-3\left[  \eta\right]  +2\right)  .\label{c3}%
\end{equation}
It is not surprising that the coefficients of the linear and quadratic
terms are the same in both models since by construction they reduce to
Eq.(\ref{lowconcentration}) at low concentrations. Note however that in
practice, when using the model of Bicerano and coworkers, Eq.(\ref{bicenario}%
), once $\left[  \eta\right]  $ is known for a given particle shape,
$\phi_{c}$ and $k_{H}$ are treated as two independent fitting
parameters when comparing to experimental data \cite{bullard}. In our model,
on the other hand, $\phi_{c}$ and $k_{H}$ are related by means
of Eq.(\ref{kH}), therefore leading to one fitting parameter only. It is
important to point out that although similar in form, expression
(\ref{viscosity1-bicenario}) has not the same divergence near $\phi_{c}%
$ as (\ref{bicenario}) since the expression in square brackets in
Eq.(\ref{viscosity1-bicenario}) diverges as $\phi\rightarrow\phi_{c}$.
However, if we make the expansion%
\begin{equation}
\eta\left(  \phi\right)  /\eta_{0}=\left(  1-\frac{\phi}{\phi_{c}}\right)
^{-\left[  \eta\right]  }\left[  1+C_{1}^{\prime}\left(  \frac{\phi}{\phi_{c}%
}\right)  +C_{2}^{\prime}\left(  \frac{\phi}{\phi_{c}}\right)  ^{2}+O\left(
\phi^{3}\right)  \right]  ,\label{viscosity1-cheng}%
\end{equation}
where%
\begin{equation}
C_{1}^{\prime}=\left[  \eta\right]  \phi_{c}-\left[  \eta\right]  ,\label{c1p}%
\end{equation}
and%
\begin{equation}
C_{2}^{\prime}=k_{H}\phi_{c}^{2}-2\left[  \eta\right]  \phi_{c}\left(  \left[
\eta\right]  -\frac{1}{2}\right)  +\left[  \eta\right]  \left(  \left[
\eta\right]  -1\right)  ,\label{c2p}%
\end{equation}
then, the expression in square brackets converges. This is an important
result meaning that the viscosity diverges near $\phi_{c}$ with an
scaling that depends on $\left[  \eta\right]  $ and therefore is not
universal but depends on the particle shape and shear rate.

\section{Illustrative calculations}

\subsection{Spheres}

The first illustrative example of the procedure is a system of hard spheres
which corresponds to take the value $\left[  \eta\right]  =5/2$ in the above
expressions which leads to the well known Einstein's result, valid for very
low concentrations%
\begin{equation}
\eta\left(  \phi\right)  =\eta_{0}\left(  1+\frac{5}{2}\phi\right)  .
\label{einstein}%
\end{equation}

Additionally to the models presented in the previous section, other
phenomenological formulas have been proposed in order to fit experiments in
the largest possible range of volume fractions. For example, Clercx and Schram
obtained the following expression for the high-frequency effective
viscosity\cite{clercx}%
\begin{equation}
\eta_{\infty}\left(  \phi\right)  =\eta_{0}\left[  1+\frac{\frac{5}{2}%
\phi+1.42\phi^{2}}{1-1.42\phi}\right]  , \label{clercx}%
\end{equation}
by considering two-particle hydrodynamic interactions only.

In order to carry out comparisons of our model with experiments and other
models it is important to notice that the value of $\phi_{c}$ in Eq.(\ref{c})
is a free parameter of the theory to be chosen in order to best fit the
experimental results. Nonetheless, this parameter can be chosen beforehand
based on physical arguments, and then used to compare with specific
experimental situations.
\begin{figure}
\center{\includegraphics[width=12cm]{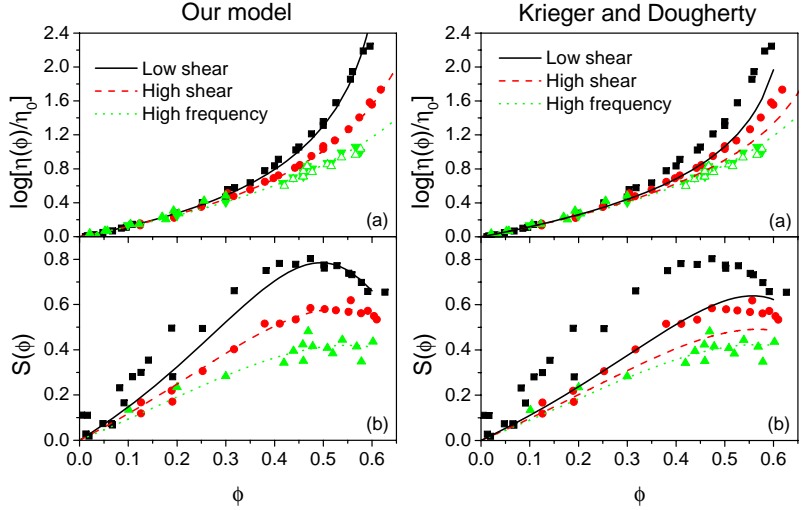}}
\caption{ 
Left panels: (a) Relative viscosity $\eta\left(  \phi\right)
/\eta_{0}$ as predicted by our model [Eq.(\ref{viscosity1})] for spheres
($\left[  \eta\right]  =5/2$) at low- and high-shear rates as well as
high-frequency as a function of the volume fraction $\phi$. The lines
correspond to the\ predictions of our model with RCP $\phi_{c}=0.63$ (upper
line), close packing at FCC $\phi_{c}=0.7404$ (middle line), and fitting
parameter $\phi_{c}=0.8678$ (lower line). The measured data are from Refs.
\cite{de Kruif} (circles) and \cite{krieger2} (triangles). Squares, Zhu
\emph{et al.} \cite{zhu}; circles, van der Werff \emph{et al.}
\cite{vanderwerff1}; triangles, Cichocki and Felderhof \cite{cichocki2}. (b)
Representation of various viscosity data as suggested by Bedeaux (Ref.
\cite{bedeaux1}). Squares, SJ18 low-shear limit (Refs. \cite{einstein} and
\cite{de Kruif}); circles, SJ18 high-shear limit (Refs. \cite{vanderwerff2}
and \cite{de Kruif}); triangles, high-frequency limit of the real part of the
complex shear viscosity (Ref.\cite{vanderwerff1}). The lines are the results
of our model with $\phi_{c}=0.63$ (upper line), $\phi_{c}=0.7404$ (medium
line), and $\phi_{c}=0.8678$ (lower line). Right panels: The same as the left
panels but the lines correspond to the model by Krieger and Dougherty
[Eq.(\ref{krieger})].
}%
\label{fig1}%
\end{figure}

This is done in the left panel of Fig. 1, where the behavior of the relative viscosity
$\eta\left(  \phi\right)  /\eta_{0}$ (Fig. 1a, left) is compared with experimental 
results of de Kruif \emph{et al.} \cite{de Kruif} and of Krieger \cite{krieger2} for low
and high-shear rates, and with the results of van der Werff \emph{et al.}
\cite{vanderwerff1} and by Zhu \emph{et al.} \cite{zhu} at high frequencies as
a function of the volume fraction $\phi$. We also plot the values obtained by
Cichocki and Felderhof \cite{cichocki2} for the high-frequency case. The upper
curve represents the prediction of Eq.(\ref{viscosity1}) with $\phi_{c}%
=0.637$, which corresponds to the random close packing (RCP) of identical
spheres. The comparison with the experimental results for low-shear rates is
excellent. The middle curve is the prediction of Eq.(\ref{viscosity1}) with
$\phi_{c}=0.7404$, which corresponds to FCC close packing. This gives again an
excellent agreement with the experimental results for the case of high-shear
rates. These results are consistent with the known fact that for low-shear
rates the spheres remain disordered while at high-shear rates the equilibrium
microstructure of the dispersion is completely destroyed and the spheres adopt
an ordered FCC configuration. The lower curve is the prediction of
Eq.(\ref{viscosity1}) with $\phi_{c}=0.8678$, which gives again an excellent
fit with the infinite-frequency data and with the values obtained by Cichocki
and Felderhof \cite{cichocki2}.

In order to account for thermodynamic interactions between the spheres,
Bedeaux proposed the following expression for the viscosity \cite{bedeaux1}%
-\cite{bedeaux2}%
\begin{equation}
\frac{\eta\left(  \phi\right)  /\eta_{0}-1}{\eta\left(  \phi\right)  /\eta
_{0}+\frac{3}{2}}=\phi\left(  1+S\left(  \phi\right)  \right)  ,
\label{bedeaux1}%
\end{equation}
where $S(\phi)$ is an unknown function of the volume fraction giving the
modification of the moment of the friction forces on the surface of a single
sphere due to the ensemble-averaged hydrodynamic interactions with the other
spheres \cite{bedeaux2}.

As discussed by Bedeaux \cite{bedeaux1}, $S\left(  \phi\right)  $ is a more
sensitive representation of the relative viscosity data since its expansion in
powers of $\phi$ converges much better. For this reason, the differences
between the predictions of the different models are more noticeable in this
representation. In Fig. 1b, left panel, we plot the function $S(\phi)$ for various
experimental results obtained in a variety of experimental situations and
compare with the values obtained with our model. We take the same values of
$\phi_{c}$ used previously for the low-shear, high-shear, and high-frequency
cases. The agreement with the experimental data is very good especially at
large values of $\phi$. Note that for small values of $\phi$ the experimental
accuracy of $S\left(  \phi\right)  $ is unsatisfactory which explains the
large scatter in the experimental points \cite{bedeaux1}.
\begin{figure}
\centerline{\includegraphics[width=12cm]{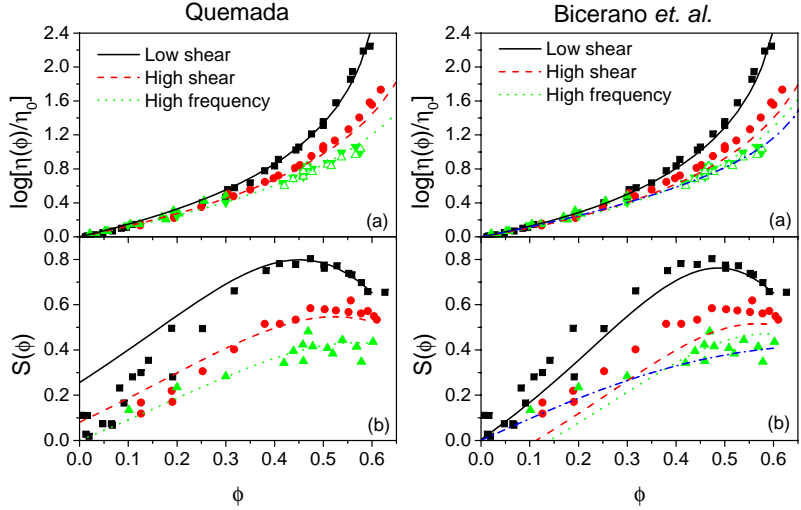}}
\caption{Left panels: The same as in Fig.1 but the lines correspond to
the model by Quemada [Eq.(\ref{quemada})]. Right panels: The same as in
Fig. 1 but the lines correspond to the model by Bicerano and
collaborators [Eq.(\ref{bicenario})]. We also show the high-frequency
result by Clercx [Eq.(\ref{clercx})] }(dash-dotted line).
\label{fig2}%
\end{figure}

As a comparative, in the right panel of Fig. 1 and in Fig. 2 we show 
the predictions obtained with the models of Krieger and Dougherty [Eq.(\ref{krieger})], 
Quemada [Eq.(\ref{quemada})], Bicerano [Eq.(\ref{bicenario})], and Clercx
[Eq.(\ref{clercx})]. In each case, the symbols (a) and (b) state for the direct 
comparison with experiments (a) and after using Bedeaux's expression (b). It is 
clear that our model matches the data at all
volume fractions better than the other models tested. 

Let us stress here that
the improvement of our model given by Eq.(\ref{viscosity1}) to the models
given by Eqs.(\ref{krieger}) and (\ref{bullard1}) is due to the use of
$\phi_{eff}$ given by Eq.(\ref{phieff}). Indeed, the model of Krieger
and Dougherty can be obtained from the dilute-limit expression $\eta\left(
\phi\right)  =\eta_{0}\left(  1+\left[  \eta\right]  \phi\right)  $ by
introducing the effective filling fraction $\phi_{eff}^{KD}\equiv\phi
/\phi_{\max}$, which is larger than $\phi$. This definition of
$\phi_{eff}^{KD}$ underestimates the available volume for the particles
at low $\phi$ (and therefore, overestimates $\phi_{eff}$) while
tends to the correct limit at high $\phi$. In order to obtain the
correct dilute limit, the overestimation of $\phi_{eff}^{KD}$, has to
be compensated by decreasing the hydrodynamic drag factor by the same constant
factor $\phi_{\max}$, that is, $\left[  \eta\right]  ^{KD}%
\equiv\left[  \eta\right]  \phi_{\max}$. Then, the dilute-limit
expression can be rewritten as%
\begin{equation}
\eta_{r}\left(  \phi\right)  =\left(  1+\left[  \eta\right]  ^{KD}\phi
_{eff}^{KD}\right)  .\label{einsteinKD}%
\end{equation}
Now, Eq. (\ref{krieger}) can be derived from (\ref{einsteinKD}) by
following the differential method used in the previous section with the
effective filling fraction $\phi_{eff}^{KD}$ instead of $\phi_{eff}$ 
and the corresponding $\left[  \eta\right]  ^{KD}$ instead of
$\left[  \eta\right]  $. The relative difference $(\phi_{eff}%
^{KD}-\phi_{eff})/\phi_{eff}$ is a decreasing function of $\phi$
that vanishes at $\phi_{\max}$ (we assumed that $\phi_{c}=\phi_{\max}$ 
in this analysis. Therefore, the overestimation of the filling
fraction in $\phi_{eff}^{KD}$, is progressively less important with
increasing $\phi$ and the constant underestimation of the hydrodynamic
drag term in Krieger and Dougherty's model cannot be compensated by %
$\phi_{eff}^{KD}$. Thus, Krieger and Dougherty's model underestimates
the viscosity of the suspension at large volume fractions, as confirmed in
Fig. 1b. The same reasoning can be applied to Bullard's model since
$1/K\equiv\phi_{c}$ plays the role of a critical concentration and
there is no mathematical distinction between Eq.(\ref{bullard1}) and the
Krieger and Dougherty equation.

In what follows, we show a direct comparison of the functional forms of our
proposal and the other models as well as a virial expansion of them. For
example, at high-frequencies, a virial expansion of Clercx and Schram's
expression, Eq.(\ref{clercx}) gives%
\begin{equation}
\eta_{\infty}\left(  \phi\right)  /\eta_{0}=1+\frac{\frac{5}{2}\phi
+1.42\phi^{2}}{1-1.42\phi}=1+\frac{5}{2}\phi+4.97\phi^{2}+7.06\phi
^{3}+O\left(  \phi^{4}\right)  ,\label{clercx-virial}%
\end{equation}
while our expression, Eq.(\ref{viscosity1}) with $\left[  \eta\right]  =5/2$
and $\phi_{c}=0.8678$, can be expressed using a Pad\'{e} approximation as%
\begin{equation}
\eta_{\infty}\left(  \phi\right)  /\eta_{0}=1+\frac{\frac{5}{2}\phi
+0.575\phi^{2}}{1-1.67\phi}=1+\frac{5}{2}\phi+4.76\phi^{2}+7.95\phi
^{3}+O\left(  \phi^{4}\right)  ,\label{viscosity1-virial}%
\end{equation}
showing that the second and third virial coefficients are very close.
Similarly, the expression by Bicerano and coworkers, Eq.(\ref{bicenario}), can
be written in the case or hard-spheres with $\phi_{c}=0.637$ as%
\begin{align}
\eta\left(  \phi\right)  /\eta_{0} &  =\left(  1-\frac{\phi}{0.637}\right)
^{-2}\left[  1-0.628\phi+0.84\phi^{2}\right]  \nonumber\\
&  =1+\frac{5}{2}\phi+6.26\phi^{2}+13.47\phi^{3}+O\left(  \phi^{4}\right)
,\label{bicenario-637}%
\end{align}
while our expression gives for the same value of $\phi_{c}$%
\begin{align}
\eta\left(  \phi\right)  /\eta_{0} &  =\left(  1-\frac{\phi}{0.637}\right)
^{-2}\left[  1-0.64\phi+0.415\phi^{2}\right]  \nonumber\\
&  =1+\frac{5}{2}\phi+5.8\phi^{2}+12.36\phi^{3}+O\left(  \phi^{4}\right)
.\label{viscosity1-637}%
\end{align}
Both expressions are very similar up to third order in $\phi.$
However, as explained before, near $\phi_{c}$ both models predict
different asymptotic relations. While Bicerano's model predicts a divergence
of the viscosity as $\left(  1-\phi/\phi_{c}\right)  ^{-2},$ our model
predicts that the viscosity diverges as $\left(  1-\phi/\phi_{c}\right)
^{-2.5},$ Eq.(\ref{viscosity1-cheng}), for the case of spherical
particles. Our prediction is in close agreement with that of Cheng et.
al.\cite{cheng}, that adapted mode coupling theories developed for supercooled
liquids and the liquid-glass transition for molecular systems to calculate the
low-shear viscosity of colloidal dispersions. Taking into account hydrodynamic
interactions, they found that near the glass transition $\phi_{g}%
=0.62$, the viscosity diverges approximately as $\left(  1-\phi
/\phi_{g}\right)  ^{-2.59}$ which is in close agreement with our model if
$\phi_{c}\simeq\phi_{g}.$
\begin{figure}
\centerline{\includegraphics[width=12cm]{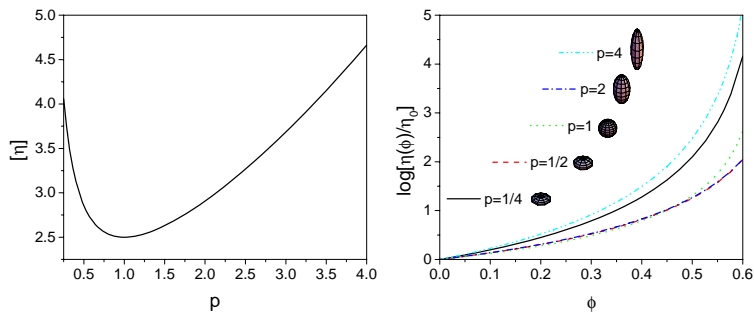}}
\caption{Left panel: Intrinsic viscosity for ellipsoids as a function of the
aspect ratio $p=b/a$. Right panel: Relative viscosity $\eta\left(
\phi\right)  /\eta_{0}$ as predicted by our model for ellipsoids with
different values of $p$. The critical packings were assumed to coincide with
the MRJ states as obtained in Refs. (\cite{donev1}) and (\cite{donev2}): for
$p=1/4$, $\phi_{c}\simeq0.64$; for $p=1/2$, $\phi_{c}\simeq0.7$; for $p=1$,
$\phi_{c}\simeq0.637$; for $p=2$, $\phi_{c}\simeq0.704$; and for $p=4$,
$\phi_{c}\simeq0.632$.}%
\label{fig3}%
\end{figure}

The proposal by Pryamitsyn and Ganesan\cite{pryamitsyn} to extend Bicerano's
expression to arbitrary shear rates by considering the packing fraction
$\phi_{c}($Pe$)$ a function of the Peclet number can also be tested by
expanding in series of $\phi$ for a large shear rate, which means to take
$\phi_{c}=0.7404$. In this case Pryamitsyn and Ganesan obtain%
\begin{align}
\eta\left(  \phi\right)  /\eta_{0} &  =\left(  1-\frac{\phi}{0.7404}\right)
^{-2}\left[  1-0.54\phi+0.622\phi^{2}\right]  \nonumber\\
&  =1+1.6\phi+3.1\phi^{2}+5.46\phi^{3}+O\left(  \phi^{4}\right)
,\label{bicenario-7404}%
\end{align}
while our model gives%
\begin{align}
\eta\left(  \phi\right)  /\eta_{0} &  =\left(  1-\frac{\phi}{0.7404}\right)
^{-2}\left[  1-0.201\phi+0.322\phi^{2}\right]  \nonumber\\
&  =1+\frac{5}{2}\phi+5.25\phi^{2}+9.94\phi^{3}+O\left(  \phi^{4}\right)
.\label{viscosity1-7404}%
\end{align}
Notice that the agreement is not very good even at first order in $\phi$ where
the extended Bicerano's expression do not agree with Einstein's expression for
low $\phi$.
\begin{figure}
\centerline{\includegraphics[width=12cm]{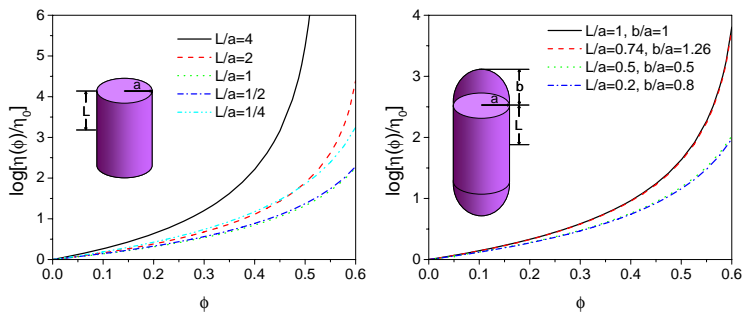}}
\caption{Left panel: Relative viscosity $\eta\left(  \phi\right)  /\eta_{0}$ as
predicted by our model for flat-ended cylinders with different aspect ratios.
Right panel: The same as left panel but for hemiellipsoidal-ended cylinders
with different aspect ratios. In both cases we assumed that the critical
packing coincides with the random packing of \ spherocylinders with the same
aspect ratio\cite{williams}.}%
\label{fig4}%
\end{figure}

\subsection{Ellipsoids}

Another case in which the behavior of highly concentrated suspensions
is of importance corresponds to suspensions of ellipsoidal particles, since
they have been widely used as models of globular proteins and also because the
hydrodynamic properties of ellipsoids are analytically known. For example, the
intrinsic viscosity of an ellipsoid is given by \cite{landau}%
\begin{align}
\left[  \eta\right]   &  =\frac{2}{5}\left(  p^{2}-1\right)  ^{2}\left\{
\frac{-\left(  4p^{2}-1\right)  B+2p^{2}+1}{3p^{2}\left(  3B+2p^{2}-5\right)
\left[  \left(  2p^{2}+1\right)  B-3\right]  }\right. \nonumber\\
&  +\frac{14}{3p^{2}\left(  3B+2p^{2}-5\right)  }+\frac{2}{\left(
p^{2}+1\right)  \left(  -3p^{2}B+p^{2}+2\right)  }\nonumber\\
&  +\left.  \frac{p^{2}-1}{p^{2}\left(  p^{2}+1\right)  \left[  \left(
2p^{2}-1\right)  B-1\right]  }\right\}  , \label{ellipsoid1}%
\end{align}
where%
\begin{align}
B  &  =p^{-1}\left(  p^{2}-1\right)  ^{-1/2}\cosh^{-1}p,\text{ \ \ \ \ \ when
\ \ \ \ \ \ }p>1,\nonumber\\
&
=1,\ \ \ \ \ \ \ \ \ \ \ \ \ \ \ \ \ \ \ \ \ \ \ \ \ \ \ \ \ \ \ \ \ \ \ \ \ \text{when
\ \ \ \ \ \ }p=1,\nonumber\\
&  =p^{-1}\left(  1-p^{2}\right)  ^{-1/2}\cos^{-1}p,\text{ \ \ \ \ \ \ \ when
\ \ \ \ \ \ }p<1. \label{B}%
\end{align}

Here, the axis ratio is defined by $p=b/a$ with $b$ the polar radius and $a$
is the equatorial radius. In the left panel of Fig. 3 we show the behavior of the intrinsic
viscosity as a function of the axis ratio $p$. As can be seen, the lower value
of the intrinsic viscosity occurs for spherical particles and increases
slightly more sharply for prolate ellipsoids than for oblate ones. In the right panel of Fig. 3
we represent the viscosity as a function of the concentration $\phi$ for
ellipsoids with different values of $p$. In all cases we have assumed that the
critical packing coincides with the so called maximally random jammed (MRJ)
state which corresponds to the least ordered among all jammed packings
\cite{torquato1}. The critical densities of simulated packings of ellipsoids
are calculated in Refs. (\cite{donev1}) and (\cite{donev2}) for different
aspect ratios and we used them in the results of Fig. 3. As can be seen, the
non-monotonic behavior of the intrinsic viscosity shown in the left panel of Fig. 3 
is reflected in the right panel of this figure.
\begin{table}[]
\begin{center}
\begin{tabular}
[c]{|c|c|c|c|c|c|} \hline
$L/a$                       & $1/4$    & $1/2$ & $1$      & $2$     & $4$      \\\hline

$\left[ \eta\right]$    & $3.92$  & $3.05$ & $2.87$ & $3.40$ & $5.27$ \\\hline

$\phi_{c} (\approx)$  & $0.67$  & $0.69$ & $0.68$ & $0.62$ & $0.53$  \\\hline

\end{tabular}
\end{center}
\vspace{0.0cm} \caption{Intrinsic viscosity $\left[  \eta\right]  $ [taken from Ref.
(\cite{zhou})] and critical packing $\phi_{c}$ [taken from Ref.
(\cite{williams})] of flat-ended circular cylinders.}
\label{table1}
\end{table}

\subsection{Cylinders}

Cylinders may be used to represent DNA molecules in certain physical
conditions and also as models for rod-shaped viruses. Flat cylinders
are encountered most often. Except in certain limit, analytical results are
not available for hydrodynamic properties of cylinders. Thus, numerical
calculations have to be used. In the limit that the total length ($2L$) goes
to zero, a flat-ended cylinder becomes a disk. The intrinsic viscosity of a
disk with radius $a$ can be found from Eq.(\ref{ellipsoid1}) by taking the
$p\rightarrow0$ limit. The result is $\left[  \eta\right]  =128a^{3}/45$. In
the case of globular cylinders for which the aspect ratio $L/a$ ranges from
$1/4$ to $4$ the hydrodynamic properties obtained numerically are listed in
Table 1, \cite{zhou}. Together with the hydrodynamic properties, in the last
column of Table 1, we show the critical packing which we assume to coincide
with the random packing of the cylinders. Actually, we are not aware of data
for the close packing of flat-ended cylinders so that we are using the results
for spherocylinders obtained in Ref. (\cite{williams}) with the same aspect
ratio. The viscosity as a function of concentration is shown in Fig. 4.

In the case of circular cylinders with hemiellipsoidal ends, the results for
the intrinsic viscosity \cite{zhou} and the critical packing are listed in
Table 2. As in the previous case, the value of the random packing listed
corresponds to spherocylinders with the same aspect ratio defined as $\left(
b+L\right)  /a$ (see inset of Fig. 5). The viscosity in this case is shown in
Fig. 5 where one can see that it is larger for the cylinders with the largest
aspect ratio.
\begin{table}[]
\begin{center}
\begin{tabular}
[c]{|c|c|c|c|c|}\hline
$L/a$                       &  $1$     &  $0.74$  & $0.5$    & $0.2$ \\\hline
$b/a$                       & $1$      & $1.26$   & $0.5 $   & $0.8 $\\\hline
$\left[\eta\right]$     & $2.944$& $2.910$ & $2.571$ & $2.512$  \\\hline
$\phi_{c}(\approx )$ & $0.62$  & $0.62$   & $0.68$   & $0.68$ \\\hline
\end{tabular}
\end{center}
\vspace{0.0cm} \caption{Intrinsic viscosity $\left[  \eta\right]  $ [taken from Ref.
(\cite{zhou})] and critical packing $\phi_{c}$ [taken from Ref.
(\cite{williams})] of hemiellipsoidal-ended circular cylinders.}
\label{table2}
\end{table}

Experimental data for the concentration and particle size-dependence of
the low-shear viscosity of isotropic rod-dispersions are discussed in
Ref.\cite{wierenga}. Four systems of stiff rods are considered: colloidal
silicaboehmite, xanthan ($\lambda=0.5$), schizophyllan ($\lambda
=0.3$) and non-Brownian PMMA-fibre. Here, $\lambda$ is the ratio
of the polymer length and the persistence length. The measured intrinsic
viscosities $\left[  \eta\right]  $ of these systems are tabulated in
Table 3. The packing density is determined by their aspect ratio. For a random
packing of thin hard rods, it was shown\cite{philipse1},\cite{philipse2} that%
\begin{equation}
\phi_{\max}\frac{L}{a}=\xi\ \ \ \ for\ \frac{L}{a}>>1, \label{phimax}%
\end{equation}
where $\xi$ is the average number of contacts experienced by a
rod. Experiments \cite{philipse1}, \cite{nardin} on random rod packing yield
$\xi/2=5.4$. This isotropic maximum packing fraction is a metastable
glass with respect to the thermodynamically more favorable nematic phase as
predicted by Onsager. The aspect ratio and the corresponding maximum packing
calculated using Eq.(\ref{phimax}) for the systems considered in Ref.
\cite{wierenga}, are also tabulated in Table 3. In Fig.5a we compare these
data to our model using as parameters the values given in Table 3. As can be
seen, poor agreement is found between the model and the experiment. A number
of reasons can be argued to explain this result. For example, the silica rods
are weakly attractive which implies a steeper viscosity-concentration curve as
compared to the rigid macromolecules xanthan and schizophyllan. Of these two
rod-like macromolecules the xanthan chain possesses higher flexibility which
may also influence the viscosity data. Additionally, since long rod-shaped
molecules "entangle" in the semidilute regime, it is expected that some
orientational correlation arises at large concentrations and our model do not
consider these effects.
\begin{table}[]
\begin{center}
\begin{tabular}
[c]{|c|c|c|c|c|}\hline
                             & Silica rods &  Schizophylian & Xanthan & PMMA-fibre\\\hline
$\left[\eta\right] $  &    $50.2$  &     $42.7$        &  $58.1$  &  $27.6$        \\\hline
$L/a$                     &    $44.0$  &     $46.0$        &  $56.0$  &  $39.8$       \\\hline
$\phi_{c}(\approx)$ &  $0.245$  &   $0.235$        & $0.193$  &  $0.271$   \\\hline
\end{tabular}
\end{center}
\vspace{0.0cm} \caption{Table 3. Intrinsic viscosity $\left[  \eta\right]  $ [taken from
Ref. (\cite{wierenga})], aspect ratio $L/a$, and critical packing
$\phi_{c}$ [obtained from Eq.(\ref{phimax})] for isotropic dispersions
of rods.} 
\label{table3}
\end{table}
\begin{figure}
\centerline{\includegraphics[width=12cm]{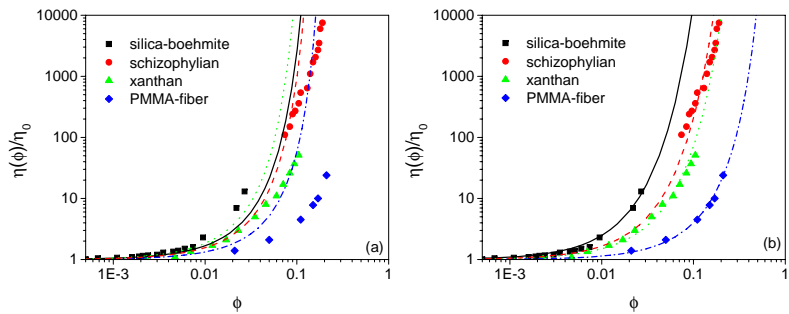}}
\caption{Left panel: Relative viscosity $\eta\left(  \phi\right)
/\eta_{0}$ as predicted by our model for isotropic rods with different
aspect ratios and intrinsic viscosities as given by Table 3. Silica rods
(solid line), schizophylian (dashed line), xanthan (dotted line), and PMMA
fibre (dash-dotted line). Right panel: The same as left panel but assuming
orientational order that leads to $\phi_{\max}\simeq0.9069$. The fitted
values of $\left[  \eta\right]  $ were $90$ for silica rods
(solid line), $50$ for schizophylian (dashed line), $40$ for
xanthan (dotted line), and $13$ for PMMA fibre (dash-dotted line). The
data sets were taken from Ref.\cite{wierenga}.
}%
\label{fig5}%
\end{figure}

To explore the possibility that the rods present some amount of
orientational correlations, we have compared our model to the experimental
data but considering that some alignment due to the flow and the correlations
is possible. In this case, the maximum packing is no longer given by
Eq.(\ref{phimax}) and we assume the opposite limit, that is, that the rods are
completely aligned. Then the maximum packing takes the value corresponding to
an hexagonal arrangement of parallel rods $\phi_{\max}\simeq0.9069$.
Then, the intrinsic viscosities for random rods as tabulated in Table 3 are no
longer useful and we take $\left[  \eta\right]  $ as a fitting
parameter. This comparison is shown in in Fig. 5b. As can be seen, a much
better agreement is obtained using these assumptions. Thus, some degree of
alignment is suggested by the model. Nonetheless, for a concentration
dependent orientational correlation, the intrinsic viscosity $\left[
\eta\right]  $ is rather a function of $\phi$ and therefore, the
values predicted by the present model may not be correct.

\subsection{Dumbbells}

A dumbbell that consists of two spheres can be used as a model for a protein
dimer or a protein consisting of two separate domains. Wakiya \cite{wakiya}
and Brenner \cite{brenner} calculated the intrinsic viscosity for dumbbells
consisting of two equal-radius spheres at various separations. Here, we will
only discuss the case when the ratio $L/a=1$, where $a$ is the radius of the
spheres and $L$ is half of their center-to-center separation. In this case,
$\left[  \eta\right]  =3.4496$.

The viscosity of dumbbells made of two fused spheres is shown in Fig. 6 and
compared with the numerical results reported by in't Veld and coworkers for
nanodimers \cite{intveld}. The closed circles represent points assuming an
effective volume fraction for particle radii adjusted to the peak onset in the
nanoparticle pair distribution function. This adjustment corrects for the
solvation shell around the nanoparticles when the solvent is treated
explicitly \cite{intveld}. The fitting parameter in this case was $\phi
_{c}\simeq0.58$, which is close to the critical packing value for the glass
transition predicted by mode coupling theory $\phi_{c}\simeq0.56$ (see Ref.
\cite{chong}).

\subsection{Other Shapes}

Notice that the intrinsic viscosity $\left[  \eta\right]  $ for
long rod-shaped particles takes values much larger as compared to the
corresponding ones for spheres. However, it is possible to have very large
values of the intrinsic viscosity without making a very extended or flat
object \cite{douglas}. An strategy to increase $\left[  \eta\right]  $
is to consider irregularly shaped particles like sponges or jack-like objects.
The intrinsic viscosities for these shapes have been calculated numerically by
finite element computations in Ref.\cite{douglas}. We use these values in Fig.
7 to compare the viscosity-concentration curves for three representative
cases: a sponge, a wire frame, a square ring, and a jack-like object.
\begin{figure}
\centerline{\includegraphics[width=10cm]{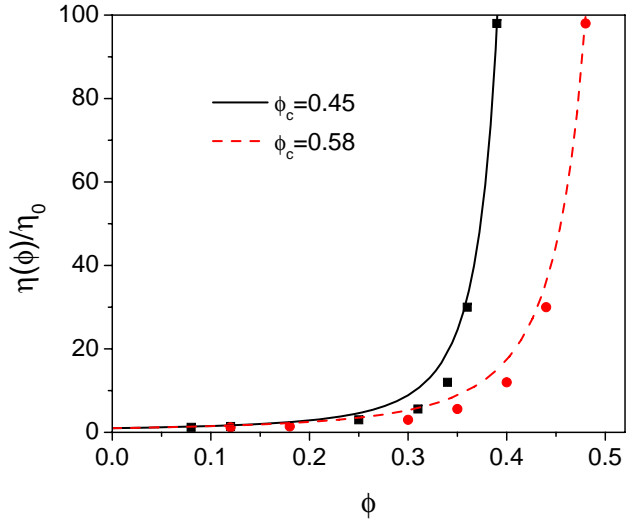}}
\caption{Relative viscosity $\eta\left(  \phi\right)  /\eta_{0}$ as predicted
by our model for a system of dumbbells. The closed squares represent points
assuming radii of the spheres $a=5\sigma$, where $\sigma$ is the size of a
Lennard-Jones solvent atom in determining $\phi$, whereas the circles denote a
volume fraction for radii adjusted to the peak onset in the nanoparticle pair
distribution function.
}%
\label{fig6}%
\end{figure}

The sponge is constructed starting with a cube in which a square channel
is cut through the center of each face, which passes completely through the
cube, as seen in Fig. 7. The parameter $m$ is taken to be the edge
length of the cutout face in units of the cube edge length. When $m$
approaches $1$ a rigid cubic wire frame is obtained. A similar
procedure is employed to construct the flat square. The jack is constructed by
poking three rectangular parallelepipeds orthogonally through a sphere. In all
the curves of Fig. 7 we assumed $\phi_{c}\simeq0.637$, the random close
packing for spheres and the values of the intrinsic viscosity given in
Ref.\cite{douglas}. As expected, the objects with larger intrinsic viscosities
have steeper curves. These examples are intended for illustrative purposes
only, since we are unaware of experimental results to compare with.

\section{Conclusions}

We have presented a simple model based on an effective-medium theory for the
calculation of the viscosity of suspensions of arbitrarily-shaped particles as
a function of particle concentration. The model considers excluded volume
interactions between the particles through an effective filling fraction
$\phi_{eff}$. This quantity introduces a universal scaling that may be used to
reduce both experimental and theoretical results to a master curve
\cite{mendoza},\cite{mendoza2} which is independent of the experimental
details or the shape of the particles.

Starting from known values and formulas for the intrinsic viscosity of the
particles, the procedure yields to analytical expressions that predict the
viscosity of the system for the whole range of concentrations. At low
filling fractions it reduces to the correct limit while at high concentrations
it diverges in a way similar to that predicted by mode coupling theories. In
contrast to other models \cite{bicenario}, our proposal contains only one
fitting parameter which corresponds to the critical packing where the
suspension loses its fluidity.
\begin{figure}
\centerline{\includegraphics[width=12cm]{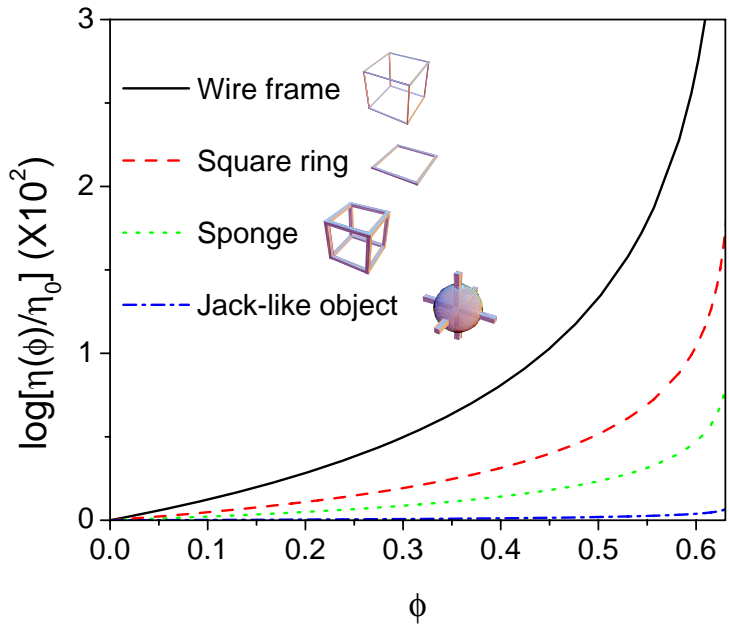}}
\caption{ Relative viscosity $\eta\left(  \phi\right)  /\eta_{0}$
as predicted by our model for a sponge ($m=23/27$, $\left[
\eta\right]  =44.7$), a cubic wire frame ($m=33/35$, $\left[
\eta\right]  =255$) a square ring ($m=23/25$, $\left[
\eta\right]  =98.7$), and a jack-like object ($\left[  \eta\right]
=3.68$). The specific proportions of the jack are as follows, if the
width of the parallelipipeds have unit length, the length has $15$
units and the sphere diameter has $9$. In all cases we assumed %
$\phi_{c}\simeq0.637$.
}%
\label{fig7}%
\end{figure}

When applied to a suspension of spherical particles, our model improves
considerably the predictions obtained using the well known Krieger and
Dougherty model and any other model tested in the whole concentration range.
We have employed our model to predict the viscosity of elliptical, and
cylindrical particles, as well as dumbbells made of fused spheres and other
complex shapes. In all cases where numerical or experimental data are
available, the agreement with the proposed model is very good. It is
convenient to emphasize that our model is not intended to describe correctly
suspensions of large fibres, since in this case orientational correlations may
exist. These correlations could introduce dependences of the intrinsic
viscosities on the concentration. In a previous work \cite{mendoza2} we have
applied the procedure to emulsions of spherical droplets with equally good
results.

Due to the importance of shape effects on the rheological behavior of
colloidal dispersions and despite that there are numerous reports for
industrial systems but fewer data for suspensions with controlled geometry, we
consider that the results presented in this article can help to provide a
valuable characterization of these systems with very promising practical applications.

\qquad{\LARGE Acknowledgements}

We thank Profs. G. S. Grest, M. K. Petersen, and P.J. in%
\'{}%
t Veld for kindly sharing with us their numerical results for a suspension of
dumbbells. This work was supported in part by Grants DGAPA IN-115010 (CIM) and
IN-102609 (ISH).

\end{document}